\documentstyle[prl,aps,amsfonts,amssymb,epsfig,floats]{revtex}
\begin{document}
\draft \preprint{}
%
%following two lines are for two column format
\twocolumn[\hsize\textwidth\columnwidth\hsize\csname
@twocolumnfalse\endcsname
\title{Fermi Surface, Surface States, and Surface Reconstruction in Sr$_2$RuO$_4$}
\author{A. Damascelli, D.H. Lu, K.M. Shen, N.P. Armitage, F. Ronning, D.L. Feng, C. Kim, and Z.-X. Shen}
\address{Department of Physics, Applied Physics and Stanford Synchrotron
        Radiation Lab., Stanford University, Stanford, CA 94305}
\author{T. Kimura, and Y. Tokura}
\address{Department of Applied Physics, The
 University of Tokyo, Tokyo 113-8656, and JRCAT, Tsukuba, 305-0046, Japan}
\author{Z.Q. Mao, and Y. Maeno}
\address{Department of Physics, Kyoto University, Kyoto 606-8502, and CREST-JST, Kawagushi,
Saitama 332-0012, Japan}

\date{Received 2 May 2000}
\maketitle
\begin{abstract}
The electronic structure of Sr$_2$RuO$_4$ is investigated by high
angular resolution ARPES at several incident photon energies. We
address the controversial issues of the Fermi surface (FS)
topology and of the van Hove singularity at the M point, showing
that a surface state and the replica of the primary FS due to
$\sqrt2\!\times\!\sqrt2$ surface reconstruction are responsible
for previous conflicting interpretations. The FS thus determined
by ARPES is consistent with the de Haas-van Alphen results, and it
provides additional information on the detailed shape of the
$\alpha$, $\beta$, and $\gamma$ sheets.
\end{abstract}
%
%74.25.Jb Electronic structure
%74.70.Ad Metals; alloys and binary compounds (including A15, Laves phases etc.)
%79.60.Bm Clean metal, semiconductor, and insulator surfaces
\vskip2pc]
\narrowtext

Angle resolved photoemission spectroscopy (ARPES) has proven
itself to be an extremely powerful tool in studying the electronic
structure of correlated electron systems. In particular, in the
case of the high-temperature superconductors, it has been very
successful in measuring the superconducting gap, determining the
symmetry of the order parameter, and characterizing the pseudo-gap
regime\cite{zx}. On the other hand one of the fundamental issues,
namely the determination of the Fermi surface (FS) topology, has
been controversial, as in the case of Bi$_2$Sr$_2$CaCu$_2$O$_8$,
raising doubts concerning the reliability of the ARPES results. A
similar controversy has also plagued the fermiology of
Sr$_2$RuO$_4$. In this context, the latter system is particularly
interesting because it can also be investigated with de Haas-van
Alphen (dHvA) experiments, contrary to the cuprates, thus
providing a direct test for the reliability of ARPES.

Whereas dHvA analysis, in agreement with LDA band-structure
calculations\cite{oguchi,singh}, indicates two electronlike FS
($\beta$, and $\gamma$) centered at the $\Gamma$ point, and a hole
pocket ($\alpha$) at the X
point\cite{mackenzie1,mackenzie2,bergemann}, early ARPES
measurements suggested a different picture: one electronlike FS
($\beta$) at the $\Gamma$ point and two hole pockets ($\gamma$,
and  $\alpha$) at the X point\cite{lu,yokoya}. The difference
comes from  the detection by ARPES of an intense, weakly
dispersive feature at the M point just below $E_F$, that was
interpreted as an extended van Hove singularity (evHs) pushed down
below $E_F$ by electron-electron correlations\cite{lu,yokoya}.
With the evHs below $E_F$, rather than above  (LDA band-structure
calculations place it 60 meV above $E_F$\cite{oguchi,singh}), the
$\gamma$ pocket is converted from electronlike to holelike. The
existence of the evHs was questioned in a later photoemission
paper\cite{puchkov}, where it was suggested that dHvA and ARPES
results could be reconciled by assuming that the feature detected
by ARPES at the M point was due to a surface state (SS). Recently,
two possible explanations were proposed: first, the evHs at the M
point could be only slightly above $E_F$ (e.g., 10 meV), so that
considerable spectral weight would be detected just below
$E_F$\cite{liebsch}. Alternatively, ARPES could be probing
ferromagnetic (FM) correlations reflected by the existence of two
different $\gamma$-FS (hole and electronlike, respectively, for
majority and minority spin direction), which escaped detection in
dHvA experiments\cite{groot}. Lastly, the surface reconstruction
as detected by LEED, which has been proposed to be indicative of a
FM surface\cite{leed}, would also complicate the ARPES data. The
resolution of this controversy is important not only for the
physics of Sr$_2$RuO$_4$ {\it per se}, but also as a reliability
test for FS's determined by ARPES, especially on those correlated
systems where photoemission is the only available probe.

In this Letter, we present a detailed investigation of the
electronic structure of Sr$_2$RuO$_4$. By varying the incident
photon energy and the temperature at which the samples were
cleaved, we confirm the SS nature of the near-$E_F$ peak detected
at the M point, and we identify an additional dispersive feature
related to the `missing' electronlike FS ($\gamma$). A complete
understanding of the data can be achieved only by recognizing the
presence of shadow bands (SB), due to the $\sqrt2\!\times\!\sqrt2$
surface reconstruction which takes place on cleaved Sr$_2$RuO$_4$
(as confirmed by LEED). Despite the surface complications, the FS
as determined by ARPES is consistent with the dHvA
results\cite{mackenzie1,mackenzie2,bergemann}, and provides
additional information on the detailed shape of the $\alpha$,
$\beta$, and $\gamma$ sheets.

ARPES data was taken at SSRL on the normal incidence monochromator
beam line equipped with a SES-200 electron analyzer in angle
resolved mode. With this configuration it is possible to
simultaneously measure multiple energy distribution curves (EDC's)
in an angular window of $\sim\!12^\circ$, obtaining
energy-momentum information not at one single $k$-point but along
an extended cut in $k$-space. The angular resolution was
0.5$^\circ$x0.3$^\circ$ [0.3$^\circ$ along the cut, corresponding
to a $k$ resolution of 1.5\% of the Brillouin zone (BZ), with 28
eV photons]. The energy resolution was 14 meV for high-symmetry
cuts and photon energy dependence, and 21 meV for the FS mappings.
Sr$_2$RuO$_4$ single crystals were oriented by Laue diffraction,
and then cleaved {\it in situ} with a base pressure better than
5x10$^{-11}$ torr. In order to compensate for the angular response
of the analyzer, EDC's in a single cut were normalized against
those measured on polycrystalline gold. Different cuts were then
normalized with respect to each other on the basis of the spectral
weight above $E_F$ integrated in both momentum and energy.
\begin{figure}[t!]
\centerline{\epsfig{figure=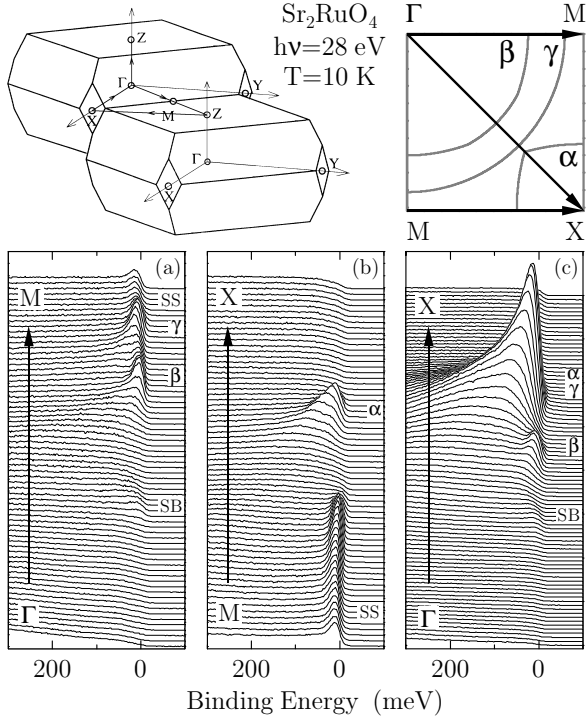,width=0.9\linewidth,clip=}}
\vspace{.1cm} \caption{ARPES spectra from Sr$_2$RuO$_4$ along the
high-symmetry lines $\Gamma$-M, M-X, and $\Gamma$-X. As shown in
the sketch of the 3D BZ, M is the midpoint along $\Gamma$-Z  and,
together with $\Gamma$ and X, defines the 2D projected zone (PZ).
In the quadrant of the 2D PZ the $\alpha$, $\beta$, and $\gamma$
sheets of FS are indicated together with the experimentally
measured cuts.} \label{edc}
\end{figure}

In order to begin the discussion of our experimental results, let
us first give an overview of the very rich photoemission spectra,
introducing all the features we will focus on throughout the
paper. Fig.\ \ref{edc} presents EDC's along the high-symmetry
directions for Sr$_2$RuO$_4$ cleaved and measured at 10 K. The
$\alpha$, $\beta$, and $\gamma$ sheets of FS expected on the basis
of LDA calculations\cite{oguchi,singh}, and dHvA
experiments\cite{mackenzie1,mackenzie2,bergemann} are indicated
together with the experimentally measured cuts in the sketch
depicting 1/4 of the 2D projected zone (PZ). All the features in
the data are labeled following the assignments which will be given
in the paper. Along $\Gamma$-M, two peaks emerge from the
background, disperse towards $E_F$, and cross it before the M
point, defining $\beta$ and $\gamma$ electronlike pockets (Fig.\
\ref{edc}a). Along M-X, a peak approaches and crosses $E_F$ before
the X point defining, in this case, the $\alpha$ hole pocket
centered at X (Fig.\ \ref{edc}b). Similar results were obtained
along $\Gamma$-X (Fig.\ \ref{edc}c): the $\beta$ pocket is clearly
resolved, while $\alpha$ and $\gamma$ crossings are almost
coincident. In addition, we identify a weak feature which shows a
dispersion opposite to the primary peaks along $\Gamma$-M and
$\Gamma$-X (SB, see below). Around the M point a sharp peak is
observed (SS), whose weak dispersion along M-X can be followed
until it crosses $E_F$ and loses intensity. The highest binding
energy along the dispersion (which is symmetric with respect to
$\Gamma$-M) is found at the M point. Because of this behavior, the
SS peak was initially associated to a sheet of FS centered at X,
and holelike in character\cite{lu}.
\begin{figure}[b!]
\centerline{\epsfig{figure=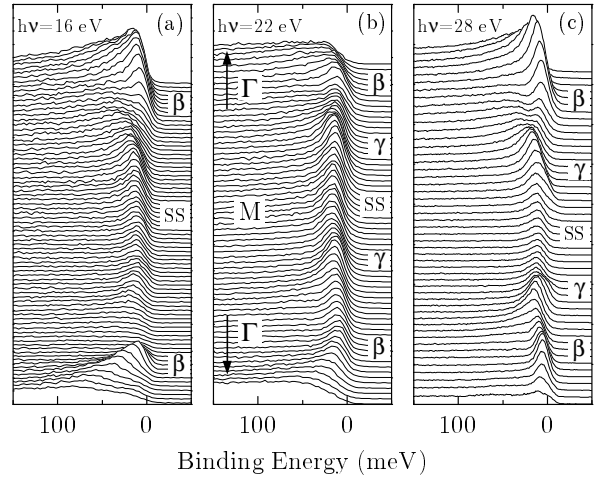,width=0.91\linewidth,clip=}}
\vspace{.1cm} \caption{ARPES spectra along $\Gamma$-M-$\Gamma$, at
three different photon energies. The cuts are centered at the M
point and extend beyond the $\gamma$ and $\beta$ FS in both first
and second zone.} \label{hv}
\end{figure}

In the following discussion, we will concentrate on the features
observed near the M point, which are relevant to the controversy
concerning the character of the $\gamma$ sheet of FS. We will show
that by working at 28 eV photon energy with sufficient momentum
resolution both the $\beta$ and $\gamma$ electronlike pockets
(predicted by LDA calculations\cite{oguchi,singh}) are clearly
resolved in the ARPES spectra. In order to address this issue, we
measured the M point region (with cuts along $\Gamma$-M-$\Gamma$)
varying the incident photon energy between 16 and 29 eV, in steps
of 1 eV. Here, we covered the location of $\beta$ and $\gamma$
pockets in both first and second zone for better illustration
(i.e., four $E_F$ crossings will be observed). From the EDC's
shown in Fig.\ \ref{hv} for 16, 22, and 28 eV, we can see that the
cross sections of SS, $\beta$, and {\it in particular} $\gamma$
exhibit a strong (and different) dependence on photon energy (note
that by working at lower photon energies the momentum resolution
increases and, in turns, the number of EDC's becomes progressively
larger in going from panel c to a). At 28\,eV, $\beta$ and
$\gamma$ crossings can be individually identified in the EDC's.
Owing to the high momentum resolution we can now follow the
dispersion of the $\gamma$ peaks until the leading edge midpoints
are located beyond $E_F$. After that, the peaks lose weight and
disappear, defining the $k_F$ vectors for the electronlike
$\gamma$ pockets. Right at $k_F$ we can resolve a double structure
which then reduces to the non dispersive feature (SS) located 11
meV below $E_F$. The difference between the 28\,eV results and
those obtained at 16 or 22\,eV is striking (the latter are
consistent with those previously reported at similar photon
energies\cite{lu,yokoya}). At low photon energies, the $\beta$
crossings are still clearly visible. On the other hand, we can
follow only the initial dispersion of the $\gamma$ peaks (now
broad and weak), before they merge with the SS feature, giving the
impression of an evHs. At 16\,eV it is impossible to identify the
$\gamma$ crossings. At 22\,eV the location of the leading edge
midpoints is at best suggestive of the presence of the $\gamma$
crossings.
\begin{figure}[t!]
\centerline{\epsfig{figure=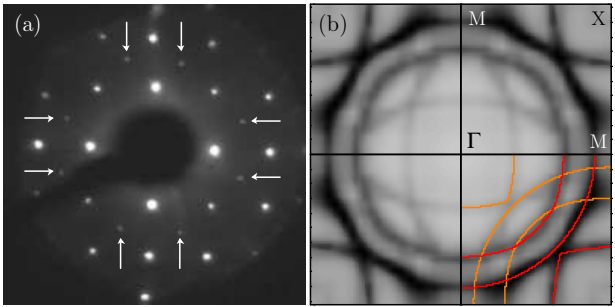,width=0.95\linewidth,clip=}}
\vspace{.1cm} \caption{(color) Panel a: LEED pattern measured at
10 K with 450 eV electrons. The arrows indicate superlattice
reflections due to $\sqrt2\!\times\!\sqrt2$ surface
reconstruction. Panel b: $E_F$ intensity map. Primary $\alpha$,
$\beta$,\,and $\gamma$ sheets of FS are marked by red lines, and
replica due to surface reconstruction by yellow lines. All data
was taken on Sr$_2$RuO$_4$ cleaved at 10 K.} \label{leed}
\end{figure}

We have showed that the $\gamma$ electron pocket had so far
escaped detection in ARPES because it is indistinguishable from
the SS feature at low photon energy and/or low angular resolution.
In order to have a full picture of the relevant issues to be
addressed, let us proceed to the discussion of the FS mapping.
Fig.\ \ref{leed}b shows the $E_F$ intensity map obtained at 28 eV
on a Sr$_2$RuO$_4$ single crystal cleaved and measured at 10 K.
The actual EDC's were taken over more than a full quadrant of the
PZ with a resolution of 0.3$^\circ$ (1$^\circ$) in the horizontal
(vertical) direction. The EDC's were then integrated over an
energy window of $\pm10$ meV about the chemical potential. The
resulting map of 73x22 points was then symmetrized with respect to
the diagonal $\Gamma$-X (to compensate for the different
resolutions along horizontal and vertical directions). The
$\alpha$, $\beta$, and $\gamma$ sheets of FS are clearly resolved,
and are marked by red lines in Fig.\ \ref{leed}b. In addition,
Fig.\ \ref{leed}b shows some unexpected features: besides the
diffuse intensity around the M point due to the presence of the SS
band, there are weak but yet well defined profiles (marked in
yellow). They can be recognized as a replica of the primary FS,
and are related to the weak SB features detected in the EDC's
along the high-symmetry lines (Fig.\ \ref{edc}a and \ref{edc}c).
This result is reminiscent of the situation found in
Bi$_2$Sr$_2$CaCu$_2$O$_8$ where similar shadow bands are possibly
related to AF correlations, or to the presence of two formula
units in the unit cell\cite{zx}. On the other hand, in
Sr$_2$RuO$_4$ the origin of the SB is completely different:
inspection with LEED reveals superlattice reflections
corresponding to a $\sqrt2\!\times\!\sqrt2$ surface reconstruction
(see Fig.\ \ref{leed}a), which is responsible for the folding of
the primary electronic structure with respect to the
($\pi$,0)-(0,$\pi$) direction. This reconstruction, which was
found on all the Sr$_2$RuO$_4$ samples, is absent in the cuprates.
Quantitative LEED analysis of the surface structure shows a
9$^\circ$ rotation of the RuO$_6$ octahedra around the surface
normal. This leads to the 45$^\circ$ rotation of the in-plane unit
cell and to the enlargement of its dimensions by
$\sqrt2\!\times\!\sqrt2$ over that of the bulk\cite{leed}. The
reconstruction, which reveals an intrinsic instability of the
cleaved surface, should be taken into account as the origin of
possible artifacts in all surface sensitive measurements.

\begin{figure*}[t!]
\centerline{\epsfig{figure=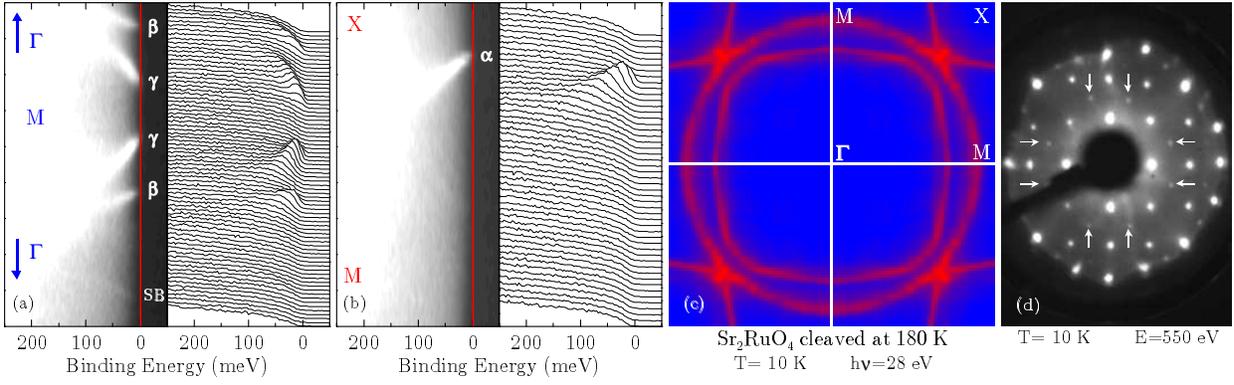,width=0.92\linewidth,clip=}}
\vspace{.1cm} \caption{(color) EDC's and intensity plot
$I(\mathbf{k},\omega)$ along $\Gamma$-M-$\Gamma$, and M-X (panel a
and b, respectively). Panel c: $E_F$ intensity map. Panel d:\,LEED
pattern recorded at the end of the FS mapping. All data was taken
at 10\,K on Sr$_2$RuO$_4$ cleaved at 180\,K.} \label{nss}
\end{figure*}

By inspecting the M point (Fig.\ \ref{leed}b), it now becomes
clear why the investigation of this $k$-space region with ARPES
has been so controversial: in addition to the weakly dispersive SS
feature (Fig.\ \ref{edc} and\ \ref{hv}), there are several sheets
of FS (primary and `folded'). At this point, the obvious question
is: what is the nature of the SS feature? It has been proposed
that it could be a surface state\cite{puchkov}, and in order to
verify this hypothesis, we investigated its sensitivity to surface
degradation by cycling the temperature between 10 and 200 K. We
observed that the SS peak is suppressed much faster than all other
features. Furthermore, by cleaving the crystals at 180 K and
immediately cooling to 10 K we suppressed the SS, affecting the
intensity of the other electronic states only weakly. A more
sizable effect is observed on the SB, confirming a certain degree
of surface degradation. However, this degradation was not too
severe, as demonstrated by the LEED pattern taken after the
measurements which still clearly shows the surface reconstruction
(Fig.\ \ref{nss}d). M-region EDC's measured at 10 K (on a sample
cleaved at 180 K), and corresponding intensity plots $I({\bf
k},\omega)$ are shown in Fig.\ \ref{nss}a and\ \ref{nss}b. No
signature of the SS is detected, and the identification of the
Fermi vectors of $\alpha$, $\beta$, and $\gamma$ pockets is now
straightforward. Performing a complete mapping on a sample cleaved
at 180 K, we obtained an extremely well defined FS (Fig.\
\ref{nss}c). With the surface slightly degraded, we expect to see
less of the relative intensity coming from SB and SS (note that
the intensity scales in Figs.\ \ref{leed}b and\ \ref{nss}c,
although not displayed, are identical). At the same time, we might
expect also the primary FS to be less well defined, which is
precisely {\it opposite} to what is observed. The FS shown in
Fig.\ \ref{nss}c is in very good agreement with LDA
calculations\cite{oguchi,singh} and dHvA
experiments\cite{mackenzie1,mackenzie2,bergemann}. The number of
electrons contained in the FS adds up to a total of 4, in
accordance with the Luttinger theorem, within an accuracy of 1\%
(as a matter of fact, for the FS determined on samples cleaved at
10 K the accuracy in the electron counting reduces to 3\% due to
the additional intensity of folded bands and surface state). As a
last remark, we can confirm that $\alpha$ and $\beta$ FS present
the nested topology which has been suggested\cite{mazin} as the
origin of the incommensurate magnetic spin fluctuations later
observed\cite{sidis} in inelastic neutron scattering experiments
at the incommensurate wave vectors ${\bf
Q}\!\approx\!(\pm2\pi/3a,\pm2\pi/3a,0)$.

Our results confirm the surface state nature of the SS peak
detected at the M point. The comparison of Figs.\ \ref{leed}b and\
\ref{nss}c suggests that a surface contribution to the total
intensity is responsible also for the less well defined FS
observed on samples cleaved at 10 K. At this point, one might
speculate that these findings are a signature of the surface
ferromagnetism (FM) recently proposed for
Sr$_2$RuO$_4$\cite{groot,leed}. In this case, two different FS's
should be expected for the two spin directions\cite{groot},
resulting in: (i) additional $E_F$-weight near M due to the
presence of a {\it holelike} $\gamma$ pocket for the majority
spin; (ii) overall momentum broadening of the FS contours because
the $\alpha$, $\beta$, and $\gamma$ sheets for the two spin
populations are slightly displaced from each other in the rest of
the BZ. Moreover, due to the surface-related nature of this
effect, it would have escaped detection in dHvA experiments. In
this scenario, a slight degradation of the surface would
significantly suppress the signal related to FM correlations, due
to the introduced disorder. The resulting FS would be
representative of the non-magnetic electronic structure of the
bulk (Fig.\ \ref{nss}c). The hypothesis of a FM surface seems
plausible because the instability of a non magnetic surface
against FM order is not only indicated by {\it ab initio}
calculations\cite{groot}, but it may also be related to the
lattice instability evidenced by the surface
reconstruction\cite{leed}. To further test this hypothesis, we
suggest spin-polarized photoemission measurements, and both linear
and nonlinear magneto-optical spectroscopy experiments [i.e.,
magneto-optical Kerr effect (MOKE), and magnetic second-harmonic
generation (MSHG), respectively].

In summary, our investigation confirms the SS nature of the weakly
dispersive feature detected at the M point (possibly a fingerprint
of a FM surface). On the basis of both ARPES and LEED, we found
that a $\sqrt2\!\times\!\sqrt2$ surface reconstruction occurs in
cleaved Sr$_2$RuO$_4$, resulting in the folding of the primary
electronic structure. Taking these findings into account, the FS
determined by ARPES is consistent with the dHvA results and
provides detailed information on the shape of $\alpha$, $\beta$,
and $\gamma$ pockets.

We gratefully acknowledge C. Bergemann, M. Braden, T. Mizokawa,
Ismail, E.W. Plummer, and A.P. Mackenzie for stimulating
discussions and many useful comments. Andrea Damascelli is
grateful to M. Picchietto and B. Top$\acute{\rm{\i}}$ for their
valuable support. SSRL is operated by the DOE office of Basic
Energy Research, Division of Chemical Sciences. The office's
division of Material Science provided support for this research.
The Stanford work is also supported by NSF grant DMR9705210 and
ONR grant N00014-98-1-0195.

\end{document}